\documentclass[conference]{IEEEtran}
\IEEEoverridecommandlockouts
\usepackage{cite}
\usepackage{amsmath,amssymb,amsfonts}
\usepackage{algorithm}
\usepackage{algorithmic}
\usepackage{algorithmic}
\usepackage{graphicx}
\usepackage{textcomp}
\usepackage{xcolor}
\usepackage{cuted}
\usepackage{caption}

\def\BibTeX{{\rm B\kern-.05em{\sc i\kern-.025em b}\kern-.08em
    T\kern-.1667em\lower.7ex\hbox{E}\kern-.125emX}}
\begin{document}

\title{
Instantaneous Risk Minimization for Secure Integrated Sensing and Communication
}

\author{\IEEEauthorblockN{Chao Ge}
\IEEEauthorblockA{\textit{Department of Electronic Engineering} \\
\textit{Tsinghua University}\\
Beijing, China \\
gechao@mail.tsinghua.edu.cn}
\and
\IEEEauthorblockN{Na Zhao}
\IEEEauthorblockA{\textit{Smart City College} \\
\textit{Beijing Union University}\\
Beijing, China \\
zhaona@buu.edu.cn}
\and
\IEEEauthorblockN{Yuan Shen}
\IEEEauthorblockA{\textit{Department of Electronic Engineering} \\
\textit{Tsinghua University}\\
Beijing, China \\
shenyuan\_ee@tsinghua.edu.cn}
}

\maketitle

\begin{abstract}
To ensure worst-case physical layer security, this paper proposes a robust beamforming framework for secure integrated sensing and communication (ISAC) systems. 
Different from conventional designs that focus on maximizing the ergodic secrecy rate, the proposed method aims to minimize instantaneous information leakage risk. 
We formulate a multi-objective optimization problem that jointly suppresses the worst-case eavesdropper signal-to-interference-plus-noise ratio (SINR), improving sensing accuracy, and ensuring the quality of service (QoS) for legitimate users. 
To address the resulting non-convex problem, we develop a hierarchical iterative algorithm, in which the outer loop refines the continuous uncertainty regions based on the updated sensing performance, and the inner loop optimizes beamforming under the refined uncertainty regions.
Theoretical analysis and simulation results demonstrate that the proposed method achieves per-transmission security guarantees with practical complexity.
\end{abstract}


\section{Introduction}
The evolution of wireless communication has enabled applications such as autonomous driving, smart factories, and digital twins, which demand stringent reliability and security~\cite{Mitev2023,Angueira2022}.
The broadcast nature of wireless channels makes them inherently vulnerable to eavesdropping~\cite{yu2021security}. 
Traditional cryptographic solutions incur high computational cost and heavy key management overhead, particularly in distributed and heterogeneous networks~\cite{Shojae2023}. Physical Layer Security (PLS) offers a lightweight and efficient alternative by exploiting intrinsic channel properties to protect confidentiality.
Despite its advantages, a long-standing bottleneck for the practical implementation of PLS is the requirement for Channel State Information (CSI) of potential eavesdroppers (Eves). In non-cooperative scenarios, where Eves conceal presence and channel characteristics, directly obtaining such information becomes infeasible, limiting the practical PLS deployment~\cite{Illi2024}.

Integrated Sensing and Communication (ISAC), a key enabler for beyond-5G and 6G networks, offers a promising breakthrough for PLS~\cite{su2024sensing,su2021}. 
ISAC aims to integrate traditional radar sensing and wireless communication functionalities into a unified hardware platform and signal processing framework~\cite{LiuISAC2022}. 
By sharing spectrum and hardware resources, ISAC not only enhances spectral and energy efficiency but also enables sensing and communication to mutually assist each other. 
In particular, ISAC systems can actively probe the environment to detect and localize non-cooperative targets. 
The native sensing capability directly aligns with the need to acquire information about potential Eves.
This transforms fully unknown threats into partially observable ones, enabling the design of robust secure beamforming strategies~\cite{Salem2025}.

Recent studies have leveraged ISAC’s dual functionality by using sensing to estimate Eve locations. 
Most existing designs adopt the ergodic secrecy rate as the performance metric, defined as the long-term average difference between the channel capacities of the legitimate user and the Eve~\cite{su2021,su2024sensing,su2025secure,Ren2022}.
While this metric quantifies the long-term average secure throughput, it hides instantaneous vulnerabilities. 
Its guarantees are statistical and cannot prevent severe information leakage during short-term channel fades favorable to the Eve. 
For security-critical applications, even a single breach can be unacceptable, making average-case guarantees insufficient.

To address this limitation, this paper introduces a novel framework for secure ISAC design that shifts the focus from maximizing the secrecy rate to ensuring \textit{worst-case instantaneous security guarantees}.
Inspired by the principles of Pointwise Maximal Leakage (PML)~\cite{PML2023}, we adopt a tractable formulation that minimizes the worst-case Eve signal-to-interference-plus-noise ratio (SINR).
The proposed framework replaces the conventional ergodic secrecy rate objective with an instantaneous security metric, thereby transitioning from average-case performance to worst-case protection. 
The communication quality-of-service (QoS) requirements are formulated as hard constraints while jointly optimizing sensing accuracy and security. 
In contrast to existing approaches that provide only statistical security assurances, our approach  considers the real-time nature of potential breaches, offering enhanced resilience against instantaneous worst-case threats.


\textit{Notations}: In this paper, vectors and matrices are denoted by bold lowercase letters (i.e., $\mathbf{x}$), and  bold uppercase letters (i.e., $\mathbf{X}$), respectively.
$\mathrm{tr}(\cdot)$ and $\mathrm{rank}(\cdot)$ denote the trace and rank operation. 
$(\cdot)^\top$, $(\cdot)^\mathrm{H}$, and $(\cdot)^*$ stand for transpose, Hermitian transpose and the complex conjugate of the matrices. 

\section{Preliminaries}
We first introduce the system model, the sensing framework to estimate potential Eve locations, and the fundamental blocks for the subsequent formulation of optimization problem.

\subsection{System model}
\begin{figure}[t]
    \centering
    \includegraphics[width= \linewidth]{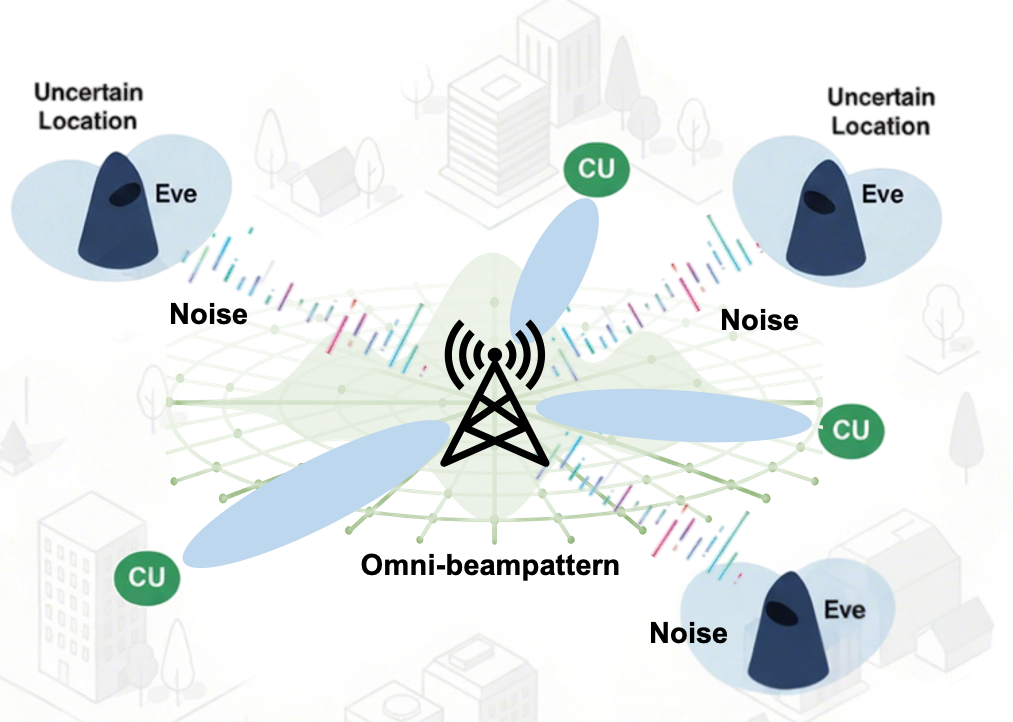}
    \captionsetup{justification=centering}
    \caption{Secure ISAC scenario with artificial noise (AN) to suppress Eve SINR.}
    \label{fig:isac}
\end{figure}

Consider an ISAC base station (BS) with $N_t$ transmit antennas and $N_r$ receive antennas, serving $I$ single-antenna communication users (CUs) in the presence of $K$ potential Eves.
The application scenario is shown in Fig. \ref{fig:isac}.

The sensing-based framework to estimate potential Eve locations is performed in two stages: active probing and parameter estimation~\cite{su2024sensing}. 
In the active probing stage, the BS transmits an omnidirectional waveform $\mathbf{X}_0$ with uniform power allocation across all directions.
Thus, the covariance matrix is given by
$\mathbf{R}_{X,0} = \frac{P_0}{N_t}\mathbf{I}_{N_t}$, where $P_0$ is the total transmit power budget.
This initial probing signal is designed to illuminate the entire angular space to search for unknown targets. 
The BS then receives the echo signals reflected from all objects within its sensing range, including both legitimate CUs and potential Eves.
The received signal is then given by
\begin{equation}\label{eq:receive}
    \mathbf{Y}_R = \sum_{k=1}^{K} \mathbf{a}(\theta_k) \beta_k \mathbf{b}^{\mathrm{H}}(\theta_k) \mathbf{X}_0 + \mathbf{Z}_R,
\end{equation}
where $\mathbf{a}(\theta_k) \in \mathbb{C}^{N_r \times 1}$, $\mathbf{b}(\theta_k) \in \mathbb{C}^{N_t \times 1}$ are the receive and transmit steering vectors for the $k$-th target, $\beta_k$ is the complex reflection coefficient of the $k$-th target, $\theta_k$ is the angle of arrival/departure for the $k$-th target, and $\mathbf{Z}_R$ is the Additive White Gaussian Noise (AWGN) with variance $\sigma_R^2$.
For a uniform linear array (ULA) with half-wavelength spacing~\cite{li2007mimo}, the steering vectors are defined as 
\begin{align*}
    &\mathbf{a}(\theta) = [e^{-j\frac{N_r-1}{2}\pi\sin\theta}, e^{-j\frac{N_r-3}{2}\pi\sin\theta}, \ldots, e^{j\frac{N_r-1}{2}\pi\sin\theta}]^{\top},\\
    &\mathbf{b}(\theta) = [e^{-j\frac{N_t-1}{2}\pi\sin\theta}, e^{-j\frac{N_t-3}{2}\pi\sin\theta}, \ldots, e^{j\frac{N_t-1}{2}\pi\sin\theta}]^{\top}.
\end{align*}

\subsection{Parameter Estimation}
In the parameter estimation stage, the received echoes are processed to estimate the parameters of unknown targets using the combined Capon and Approximate Maximum Likelihood (CAML) technique~\cite{su2024sensing}.
This approach first applies the Capon method to obtain initial estimates of the directions of all reflected paths, denoted as $\hat{\boldsymbol{\theta}}:=(\hat{\theta}_1,\ldots,\hat{\theta}_K )$.
Similar to \eqref{eq:receive}, we have $\mathbf{Y}=\mathbf{A}^*(\hat{\boldsymbol{\theta}})\mathrm{diag}\left[\beta(\hat{\theta}_1),\ldots,\beta(\hat{\theta}_K)\right]\mathbf{B^\top(\hat{\boldsymbol{\theta}})}\mathbf{X}+\tilde{\mathbf{Z}}$, where $\tilde{\mathbf{Z}}$ denotes the residual term.
Then, the approximate maximum likelihood algorithm is employed to estimate the corresponding complex amplitudes:
\begin{multline*}
    \boldsymbol{\beta} = \frac{1}{L}\left[\left(\mathbf{A}^{\mathrm{H}}\mathbf{T}^{-1}\mathbf{A}\right) \odot \left(\mathbf{B}^{\mathrm{H}}\hat{\mathbf{R}}_{X,0}^*\mathbf{B}\right)\right]^{-1} \\
    \cdot \mathrm{vecd}\left(\mathbf{A}^{\mathrm{H}}\mathbf{T}^{-1}\mathbf{Y}_R\mathbf{X}_0^{\mathrm{H}}\mathbf{B}^*\right),
\end{multline*}
where $L$ is the number of time-domain snapshots, 
$\mathbf{A} = \left[\mathbf{a}(\hat{\theta}_1), \ldots, \mathbf{a}(\hat{\theta}_K)\right]$, $\mathbf{B} = \left[\mathbf{b}(\hat{\theta}_1), \ldots, \mathbf{b}(\hat{\theta}_K)\right]$,
$\mathbf{T} = L\hat{\mathbf{R}} - \frac{1}{L}\mathbf{Y}_R\mathbf{X}_0^{\mathrm{H}}\mathbf{B}^*(\mathbf{B}^{\top}\hat{\mathbf{R}}_{X,0}\mathbf{B}^*)^{-1}\mathbf{B}^{\top}\mathbf{X}_0\mathbf{Y}_R^{\mathrm{H}}$, $\hat{\mathbf{R}}$ is the sample covariance of the observed data samples and
$\hat{\mathbf{R}}=\frac{1}{L}\mathbf{Y}\mathbf{Y}^{\mathrm{H}}$.
The operator
$\odot$ denotes element-wise multiplication, and $\mathrm{vecd}(\cdot)$ extracts the diagonal elements as a column vector.
It is assumed that the locations of the CUs are known to the BS, then their corresponding echoes can be identified and excluded.
The remaining parameter estimates thus correspond to non-cooperative targets, which are subsequently treated as potential Eves.
The estimates for the reflection coefficients are bounded by the Cramér-Rao Bound (CRB), a fundamental lower bound on the accuracy of the parameter estimates.

\subsection{Cramér-Rao Bound (CRB)}
To compute the CRB for parameter estimation, we first derive the Fisher Information Matrix (FIM).
Let $\boldsymbol{\eta} := \left[\theta_1, \ldots, \theta_K, \mathrm{Re}(\beta_1), \mathrm{Im}(\beta_1), \ldots, \mathrm{Re}(\beta_K), \mathrm{Im}(\beta_K)\right]^{\top}$ denote the parameter vector to be estimated, where $\mathrm{Re}(\cdot)$ and  $\mathrm{Im}(\cdot)$ denote the real and imaginary parts, respectively.
The first-order derivatives of the steering vectors with respect to the angular parameter are given by
$$
\dot{\mathbf{A}} = \left[\frac{\partial\mathbf{a}(\theta_1)}{\partial\theta_1}, \ldots, \frac{\partial\mathbf{a}(\theta_K)}{\partial\theta_K}\right],
~\dot{\mathbf{B}} = \left[\frac{\partial\mathbf{b}(\theta_1)}{\partial\theta_1}, \ldots, \frac{\partial\mathbf{b}(\theta_K)}{\partial\theta_K}\right].
$$
Let $\mathbf{C}_A:=\mathbf{A}^{\mathrm{H}}\mathbf{A}$ denote the receive side Gramian matrix, and $\mathbf{C}_B:=\mathbf{B}^{\mathrm{H}}\mathbf{R}_{X,0}^*\mathbf{B}$ denote the transmit side weighted Gramian matrix.

To quantify the correlation between the steering vectors and their first-order derivatives with respect to the angular parameters, let $\mathbf{D}_A:=\dot{\mathbf{A}}^{\mathrm{H}}\mathbf{A}$ and $\mathbf{D}_B:=\dot{\mathbf{B}}^{\mathrm{H}}\mathbf{R}_{X,0}^*\mathbf{B}$.
Analogous to the standard Gramian matrices, let $\mathbf{E}_A:=\dot{\mathbf{A}}^{\mathrm{H}}\dot{\mathbf{A}}$ and $\mathbf{E}_B:=\dot{\mathbf{B}}^{\mathrm{H}}\mathbf{R}_{X,0}^*\dot{\mathbf{B}}$.
Denote $\boldsymbol{\Lambda} := \mathrm{diag}(\beta_1.\ldots,\beta_K)$.
Therefore, the FIM is given by
$$\mathbf{J} = \frac{2L}{\sigma_R^2}\begin{bmatrix}
\mathrm{Re}(\mathbf{J}_{11}) & \mathrm{Re}(\mathbf{J}_{12}) & -\mathrm{Im}(\mathbf{J}_{12}) \\
\mathrm{Re}^{\top}(\mathbf{J}_{12}) & \mathrm{Re}(\mathbf{J}_{22}) & -\mathrm{Im}(\mathbf{J}_{22}) \\
-\mathrm{Im}^{\top}(\mathbf{J}_{12}) & -\mathrm{Im}^{\top}(\mathbf{J}_{22}) & \mathrm{Re}(\mathbf{J}_{22})
\end{bmatrix},$$
where the submatrices are defined as \eqref{eq:j11} - \eqref{eq:j22}.
\begin{align}
\mathbf{J}_{11} &= \mathbf{E}_A \odot \left(\boldsymbol{\Lambda}^*\mathbf{C}_B\boldsymbol{\Lambda}\right)  + \mathbf{D}_A \odot \left(\boldsymbol{\Lambda}^*\mathbf{D}_B^{\mathrm{H}}\boldsymbol{\Lambda}\right) \nonumber\\&~~~+ \mathbf{D}_A^{\mathrm{H}} \odot \left(\boldsymbol{\Lambda}^*\mathbf{D}_B \boldsymbol{\Lambda}\right) +  \mathbf{C}_A \odot \left(\boldsymbol{\Lambda}^*\mathbf{E}_B\boldsymbol{\Lambda}\right)  \label{eq:j11}\\
\mathbf{J}_{12} &= \boldsymbol{\Lambda}^* \odot \left( \mathbf{D}_A \odot \mathbf{C}_B  +  \mathbf{C}_A \odot \mathbf{D}_B  \right) \label{eq:j12}\\
\mathbf{J}_{22} &=  \mathbf{C}_A \odot \mathbf{C}_B \label{eq:j22} 
\end{align}
Then, the corresponding CRB matrix is given by
    $\mathrm{CRB}(\boldsymbol{\eta}) = \mathbf{J}^{-1}$.
For the angular parameters, the diagonal entries of the inverse FIM yield
\begin{equation}\label{eq:CRB-k}
    \mathrm{CRB}(\theta_k) = \left[\mathbf{J}^{-1}\right]_{k,k},~k=1,\ldots,K.
\end{equation}
Assuming Gaussian estimation errors, the angular uncertainty region for each detected Eve is defined as
\begin{equation}\label{eq:eve}
    \Xi_k^{(0)} = \left[\hat{\theta}_k - \xi\sqrt{\mathrm{CRB}(\theta_k)},\ \hat{\theta}_k + \xi\sqrt{\mathrm{CRB}(\theta_k)}\right],
\end{equation}
where $\xi$ is the confidence interval parameter (e.g., $\xi = 3$ corresponds to a 99.7\% confidence level).
The root mean square error (RMSE) of angle estimation is thus bounded by $\mathrm{RMSE}(\theta_k) \geq \sqrt{\mathrm{CRB}(\theta_k)}$.
The obtained angular uncertainty regions $\{\Xi_k^{(0)}\}$ serve as the foundation for the subsequent optimization problem.

\section{Problem Formulation}
In this section, we formulate a novel optimization problem for secure ISAC systems. 
Unlike conventional designs based on the ergodic secrecy rate, we adopt a robust formulation that ensures \textit{worst-case security guarantees}. 
The objective is to jointly enhance sensing accuracy and minimize the instantaneous worst-case information leakage risk, while guaranteeing the QoS requirements for all legitimate CUs.

\subsection{Performance Metrics}
Following~\cite{su2024sensing},
the BS transmits a dual-functional signal with a total power budget of $P_0$. The optimization variables include the beamforming covariance matrices, $\tilde{\mathbf{W}}_i = \mathbf{w}_i\mathbf{w}^{H}_i \in \mathbb{C}^{N_t\times N_t}$, for $i=1,\ldots,I$, and the AN covariance matrix $\mathbf{R}_N \in \mathbb{C}^{N_t\times N_t}$. 

\subsubsection{Sensing Performance}
The sensing performance is characterized by the FIM $\mathbf{J}$.
A larger $|\mathbf{J}|$ corresponds to a smaller CRB and thus higher estimation accuracy. 
For numerical stability, we adopt $\log |\mathbf{J}|$ as the sensing objective~\cite{li2007mimo,boyd2004convex}.

\subsubsection{Security Performance}
Conventional ISAC security designs are based on secrecy rate, which captures only long-term average performance. We instead adopt an instantaneous metric that minimizes worst-case information leakage per transmission. Inspired by PML~\cite{PML2023}, we leverage the fact that an adversary’s symbol distinguishability increases monotonically with SINR in AWGN channels. Thus, we define the security objective as minimizing the worst-case SINR across all Eves and angular uncertainties, denoted as $\mathrm{SINR}_{k,i}^E(\theta)$.

However, directly minimizing $\max_{k,i,\theta} \mathrm{SINR}_{k,i}^E(\theta)$ leads to a non-smooth min–max structure. To obtain a tractable form, we adopt the \textit{epigraph reformulation} by introducing an auxiliary variable $t \geq 0$, which represents the upper bound on the worst-case SINR at any Eve.
This smooth reformulation transforms the original intractable objective into a constrained optimization problem with a linear objective.

\subsection{Proposed Optimization Problem}
We now integrate sensing and security objectives into a unified framework which jointly maximizes sensing accuracy and minimizes worst-case instantaneous leakage, thereby providing a more realistic and robust foundation for security-critical ISAC systems.

Let $\rho \in [0,1]$ denote the weighting factor between sensing and security goals. The objective function is formulated as
\begin{equation}\label{eq:opti}
    \max_{\tilde{\mathbf{W}}_i,\mathbf{R}_N,t} \rho \cdot \frac{\log|\mathbf{J}|}{\log|\mathbf{J}|_{\mathrm{norm}}} - (1-\rho) \cdot \frac{t}{t_{\mathrm{norm}}},
\end{equation}
where $\log|\mathbf{J}|_{\mathrm{norm}}$ and $t_{\mathrm{norm}}$ are normalization constants, ensuring comparable scales between the two goals.
The optimization subjects to the following constraints:

\textbf{(C1) Instantaneous Security:} The SINR for any Eve $k$ eavesdropping on any CU $i$'s signal must be upper-bounded by $t$ across the angular uncertainty region $\Xi_k$, by (9) of \cite{su2024sensing}, 
\begin{multline}\label{eq:c1}
    \mathrm{SINR}_{k,i}^E(\theta) \\
    =\frac{|\alpha_k|^2 \mathbf{b}^{\mathrm{H}}(\theta)\tilde{\mathbf{W}}_i \mathbf{b}(\theta)}{|\alpha_k|^2 \mathbf{b}^{\mathrm{H}}(\theta)\left(\sum_{m \neq i}\tilde{\mathbf{W}}_m + \mathbf{R}_N\right)\mathbf{b}(\theta) + \sigma_0^2}  \leq t, \\
    \forall i,k,\forall \theta \in \Xi_k,
\end{multline}
where $\alpha_k$ represents the channel gain, $\mathbf{b}(\theta) $ is the steering vector at angle $\theta$, and $\sigma_0^2$ is the noise power at the Eves.

\textbf{(C2) CU QoS Requirements:} The received SINR of each legitimate CU $i$ must satisfy a minimum required threshold $\gamma_i$.
This threshold corresponds to the minimum communication rate required by CU $i$, given by $R_i^{\min} =\log\left( 1+\gamma_i\right)$.
Hence, 
\begin{equation}\label{eq:c2}
    \frac{\mathrm{tr}\left(\tilde{\mathbf{H}}_i \tilde{\mathbf{W}}_i \right)
    }{\sum_{m\neq i}\mathrm{tr}\left(\tilde{\mathbf{H}}_i \tilde{\mathbf{W}}_m\right) + \mathrm{tr}\left(\tilde{\mathbf{H}}_i \mathbf{R}_N\right) + \sigma_C^2} \geq \gamma_i,~\forall i,
\end{equation}
where $\tilde{\mathbf{H}}_i = \mathbf{h}_i \mathbf{h}_i^{\mathrm{H}}$ is the channel covariance matrix for CU $i$, and $\sigma_C^2$ is the noise power at the legitimate CUs.

\textbf{(C3) Sensing Beampattern:} To ensure effective sensing functionality, the beampattern of the total transmitted signal, defined by $\mathbf{R}_X =\sum_i \tilde{\mathbf{W}}_i + \mathbf{R}_N$, must satisfy mainlobe and sidelobe power constraints~\cite{su2024sensing}.
The mainlobe constraint focuses power toward potential targets and the sidelobe constraint suppresses interference:
\begin{subequations}
\begin{align}
&\mathbf{b}^{\mathrm{H}}(\theta) \mathbf{R}_X \mathbf{b}(\theta) \geq (1-\alpha)\mathbf{b}^{\mathrm{H}}(\hat{\theta}_{k})\mathbf{R}_X \mathbf{b}(\hat{\theta}_{k}), \forall \theta \in \Xi_k, \forall k, \label{eq:c3a}\\
&\mathbf{b}^{\mathrm{H}}(\theta) \mathbf{R}_X \mathbf{b}(\theta) \leq \gamma_s , \forall \theta \in \Psi_k, \forall k, \label{eq:c3b}
\end{align}
\end{subequations}
where $\alpha \in [0,1]$ controls the allowable power ripple within the mainlobe,
$\gamma_s$ denotes the given threshold to constrain the power of the sidelobe, and $\Psi_k$ denotes the sidelobe region.

\textbf{(C4) Total Power:} The total transmit power is limited by the power budget $P_0$, that is,
\begin{equation}\label{eq:c4}
    \mathrm{tr}\left( \sum_{i=1}^I \tilde{\mathbf{W}}_i + \mathbf{R}_N\right) \leq P_0.
\end{equation}

\textbf{(C5) Matrix Property:} The covariance matrices must be positive semidefinite, and the beamforming matrices must be rank-one. Therefore,
\begin{equation}\label{eq:c5}
    \tilde{\mathbf{W}}_i \succeq \mathbf{0}, \mathbf{R}_N \succeq \mathbf{0}, \mathrm{rank} \left( \tilde{\mathbf{W}}_i\right) =1, \forall i.
\end{equation} 


The normalization constants are determined by finding the upper bound of each metric by solving two single-objective optimization problems. 
These values can be precomputed by solving the respective single-objective problems using the algorithmic framework outlined in Section \ref{sec:solution}.

\section{Solution Algorithm}\label{sec:solution}
The formulated problem \eqref{eq:opti} - \eqref{eq:c5} is highly non-convex and computationally challenging.
The non-convexity arises from the semi-infinite constraints (C1) and (C3), the bilinear terms in the security constraint (C1), and the rank-one constraint (C5). Moreover, the semi-infinite constraints (C1) and (C3), defined over continuous angular uncertainty regions, further increase the complexity.

To address these challenges, we develop a hierarchical iterative optimization framework. The outer loop progressively refines the angular uncertainty regions based on updated sensing information, while the inner loop solves the resulting subproblems by discretizing the continuous regions and employing convex approximation techniques. This decomposition enables tractable optimization while retaining robustness against angular uncertainty.


\subsection{Outer Loop: Iterative Uncertainty Refinement}
The outer loop progressively improves the estimation of the Eves' locations by updating their angular uncertainty regions according to the CRB. At each iteration, the beamforming solution obtained under the current uncertainty sets is used to update the FIM, from which refined uncertainty bounds are derived. This establishes a feedback mechanism between sensing and security optimization.

\begin{algorithm}[H]
\caption{Iterative Uncertainty Refinement}
\begin{algorithmic}[1]
\STATE \textbf{Initialize} uncertainty regions $\Xi_k^{(0)}$ from initial sensing
\FOR{iteration $r = 1, 2, \ldots$ until convergence}
    \STATE \textbf{Solve} problem \eqref{eq:opti} - \eqref{eq:c5} within $\Xi_k^{(r-1)}$ by inner loop to obtain $\{\tilde{\mathbf{W}}_i^\star, \mathbf{R}_N^\star, t^\star\}$ 
    \STATE \textbf{Update} FIM $\mathbf{J}^{(r)} \gets \mathbf{J}(\mathbf{R}_X^\star)$ with $\mathbf{R}_X^\star = \sum_i \tilde{\mathbf{W}}_i^\star + \mathbf{R}_N^\star$
    \STATE \textbf{Compute} CRB: $\mathrm{CRB}(\theta_k) = [\mathbf{J}^{(r)^{-1}}]_{k,k}$
    \STATE \textbf{Update} uncertainty regions: 
    $$
    \Xi_k^{(r)} = \left[\hat{\theta}_k - \xi\sqrt{\mathrm{CRB}(\theta_k)},\ \hat{\theta}_k + \xi\sqrt{\mathrm{CRB}(\theta_k)}\right]
    $$
\ENDFOR
\STATE \textbf{Return} final beamforming and AN covariance matrices
\end{algorithmic}
\end{algorithm}

\subsection{Inner Loop: Solving problem \eqref{eq:opti}}
Given the current uncertainty regions $\Xi_k$, the inner loop solves problem \eqref{eq:opti}-\eqref{eq:c5} via discretization, semidefinite relaxation (SDR), and sequential convex approximation (SCA).

\textbf{Step 1: Discretization and SDR.}
First, we discretize the continuous angular regions $\Xi_k$ and $\Psi_k$ into finite sets:
\begin{align*}
    &\Xi_k \rightarrow \hat{\Xi}_k = \{\theta_{k,1}, \theta_{k,2}, \ldots, \theta_{k,L_k}\},\\
    &\Psi_k \rightarrow \hat{\Psi}_k = \{\psi_{k,1}, \psi_{k,2}, \ldots, \psi_{k,M_k}\},
\end{align*}
where $L_k$ and $M_k$ denote the respective numbers of discretization points. This step transforms the semi-infinite constraints (C1) and (C3) into a finite set of constraints.
It should be noted that approximation accuracy depends on the discretization resolution. 
Finer discretization improves robustness by more tightly enforcing the constraints, at the expense of higher computational cost.

The rank-one constraints on $\tilde{\mathbf{W}}_i$ are relaxed via SDR method: $\tilde{\mathbf{W}}_i \succeq \mathbf{0}$. 
If the optimal solution of the relaxed problem satisfies the rank-one property, it is also an optimal solution of the original problem. Otherwise, a suitable rank-one solution must be recovered.

\textbf{Step 2: Sequential Convex Approximation.}
The relaxed problem remains non-convex due to the bilinear terms in the discretized security constraints (C1). We solve it via an iterative SCA procedure. At each inner loop iteration $n=1,2,\ldots$, we solve a convex approximated problem formulated at the solution point from the previous iteration, $\{ \tilde{\mathbf{W}}_i^{(n-1)},\mathbf{R}_N^{(n-1)} , t^{(n-1)}\}$.
For the security constraints (C1), we first reformulate them as
\begin{multline*}
    |\alpha_k|^2 \mathbf{b}^{\mathrm{H}}(\theta_{k,j})\tilde{\mathbf{W}}_i \mathbf{b}(\theta_{k,j}) \\
    \leq t\left(|\alpha_k|^2 \mathbf{b}^{\mathrm{H}}(\theta_{k,j})\left(\sum_{m \neq i}\tilde{\mathbf{W}}_m + \mathbf{R}_N\right)\mathbf{b}(\theta_{k,j}) + \sigma_0^2 \right),
\end{multline*}
for all discretized angle $\theta_{k,j} \in \hat{\Xi}_k$.
This form contains bilinear terms due to the product of $t$ and the interference terms. 
To handle the bilinear terms, we apply SCA by replacing one of the variables in the product with its value from the previous iteration. This linearizes the constraint within the current subproblem.
Specifically, 
let $s_{k,i,j}:=|\alpha_k|^2 \mathbf{b}^{\mathrm{H}}(\theta_{k,j})\left(\sum_{m \neq i}\tilde{\mathbf{W}}_m + \mathbf{R}_N\right)\mathbf{b}(\theta_{k,j}) + \sigma_0^2$.
Thus, the bilinear term $t \cdot s_{k,i,j}$ can be convexified using the following approximation 
$$
t \cdot s_{k,i,j} \approx t^{(n-1)} \cdot s_{k,i,j}+  s_{k,i,j}^{(n-1)}\left( t -t^{(n-1)} \right).
$$
This results in a constraint that is linear in the optimization variables for the current iteration.

For the QoS constraints (C2), \eqref{eq:c2} can be equivalently rewritten as the affine inequality
\begin{equation*}
\mathrm{tr}(\tilde{\mathbf{H}}_i \tilde{\mathbf{W}}_i)
- \gamma_i\sum_{m\neq i}\mathrm{tr}(\tilde{\mathbf{H}}_i \tilde{\mathbf{W}}_m)
- \gamma_i\,\mathrm{tr}(\tilde{\mathbf{H}}_i\mathbf{R}_N)
\ge \gamma_i\,\sigma_C^2,
\end{equation*}
which is a convex affine constraint and can be directly included in the SDP subproblem.
Therefore, at each SCA iteration, the approximated problem reduces to a standard SDP with a concave objective subject to linear and SDP constraints that can be solved efficiently by interior-point methods.
The SCA procedure generates a sequence of solutions that is guaranteed to converge to a stationary point of the original problem.
Upon convergence, the algorithm outputs the stationary solution
$\{\tilde{\mathbf{W}}_i^\star, \mathbf{R}_N^\star, t^\star\}$.
The detailed algorithm is presented as follows.

\begin{algorithm}[H]
\caption{Sequential Convex Approximation}
\begin{algorithmic}[1]
\STATE \textbf{Initialize}  with a feasible point $\{\tilde{\mathbf{W}}_i^{(0)}, \mathbf{R}_N^{(0)}, t^{(0)}\}$
\FOR{iteration $n = 1, 2, \ldots$ until convergence}
    \STATE \textbf{Convexify SINR constraints}: Approximate bilinear terms in (C1) using first-order expansions
    \STATE \textbf{Solve SDP}: Optimize the convexified subproblem using interior-point methods
    \STATE \textbf{Update} $\{\tilde{\mathbf{W}}_i^{(n)}, \mathbf{R}_N^{(n)}, t^{(n)}\}$ with the obtained solution
    \STATE \textbf{Check convergence}; if satisfied, terminate
\ENDFOR
\STATE \textbf{Return} $\{\tilde{\mathbf{W}}_i^\star, \mathbf{R}_N^\star, t^\star\}$ 
\end{algorithmic}
\end{algorithm}

\textbf{Step 3: Rank-One Solution Recovery.}
Upon convergence of the SCA iterations, the obtained covariance matrices ${\tilde{\mathbf{W}}_i^\star}$ may not necessarily be rank-one. Since rank-one beamforming matrices are required for practical implementation, we adopt the standard Gaussian randomization method to recover feasible beamforming vectors.
Specifically, for each CU $i$, 
we generate a set of $N_{\mathrm{rand}}$ independent candidate beamforming vectors sampled as $\mathbf{v}_{i,\ell} \sim \mathcal{CN}(\mathbf{0},\tilde{\mathbf{W}}_i^\star)$, where $\ell = 1,\ldots,N_{\mathrm{rand}}$. 
Each candidate vector is then used to construct a rank-one approximation $\hat{\mathbf{W}}_{i,\ell} = \mathbf{v}_{i,\ell}\mathbf{v}_{i,\ell}^{\mathrm{H}}$.
To ensure feasibility under the transmit power budget, all candidate beamformers are uniformly scaled such that the total transmit power constraint is satisfied with equality.
Among the $N_{\mathrm{rand}}$ randomized candidates, we select the set of beamforming matrices ${\hat{\mathbf{W}}_i^\star}$ that maximizes the original objective function while satisfying all constraints. This procedure guarantees that the final solution is both feasible and practically implementable, while closely approximating the optimal value achieved by the relaxed SDP solution.

\subsection{Convergence Analysis}
We now analyze the convergence behavior of both the inner and outer iterative loops, establishing the stability of the proposed algorithm.

\subsubsection{Convergence of the inner loop}
For the inner loop, 
we employ an SCA procedure to handle the non-convexity arising from the bilinear terms in the security constraints (C1).
Consequently, at each iteration $n$, 
the non-convex constraints are replaced by their first-order Taylor approximations around the previous point  $\{ \tilde{\mathbf{W}}_i^{(n-1)}, \mathbf{R}_N^{(n-1)}, t^{(n-1)} \}$, resulting in a standard SDP subproblem. This subproblem involves maximizing a concave objective function, i.e., the original objective \eqref{eq:opti}, over a convex set defined by the approximated constraints.
The feasible set of the original problem is compact due to the bounded transmit power constraint (C4) and the positive semidefinite constraints (C5).
The SCA method generates a sequence of solutions $\{ \tilde{\mathbf{W}}_i^{(n)}, \mathbf{R}_N^{(n)}, t^{(n)} \}$.
By standard results in the theory of SCA methods, this sequence is guaranteed to converge to a stationary solution, i.e., a KKT point of the discretized and relaxed problem.


\subsubsection{Convergence of the outer loop}
The outer loop refines the angular uncertainty regions $\Xi_k^{(r)}$. 
In each iteration $r$, improves sensing accuracy by maximizing $\log |\mathbf{J}|$ reduces the CRB, $\mathrm{CRB}(\theta_k) = [\mathbf{J}^{(r)^{-1}}]_{k,k}$,
which in turn decreases the width of the uncertainty interval, $\Xi_k^{(r)}$.
Since $|\mathbf{J}|$ increases monotonically across iterations, the uncertainty widths decrease monotonically. Furthermore, the widths are bounded below by zero, which ensures that the sequence of region sizes ${|\Xi_k^{(r)}|}$ converges. Therefore, the outer loop is guaranteed to stabilize, yielding refined angular uncertainty sets.




\section{Simulation Results}
In this section, we provide numerical results to validate the effectiveness of our proposed hierarchical algorithm for secure ISAC. We evaluate the convergence behavior, the iterations of key performance metrics, and the spatial characteristics of the optimized beamforming solution.

\subsection{Simulation Setup}
We consider a multiple-input multiple-output (MIMO) ISAC system where a base station BS equipped with a ULA serves multiple CUs in the presence of a potential Eve. The channel vectors for the CUs are modeled as independent and identically distributed complex Gaussian entries. 
The core optimization subproblem in each iteration is solved using the CVX modeling framework with the SDPT3 solver~\cite{grant2008cvx}. 
Unless stated otherwise, the simulation parameters are summarized as follows.
The BS is equipped with $N_t = 8$ transmit antennas and serves $I = 2$ CUs. 
The total transmit power budget is set to $P_0=20~\mathrm{dBm}$.
The noise power at both the CUs and the eavesdropper is set to $\sigma_C^2 = \sigma_0^2 = -60$ dBm. The minimum SINR requirements for the two CUs are set to $\gamma_1=\gamma_2 = 10$ dB.
The confidence interval parameter is set to $\xi=3$.


\subsection{Performance Evaluation}
We first evaluate the convergence behavior and the evolution of key performance metrics for the proposed hierarchical algorithm. For comparison, we benchmark our algorithm against the secrecy rate (SR)-proxy optimization method from \cite{su2024sensing}.
\begin{figure*}[t]
\centering
\begin{minipage}[b]{0.32\linewidth}
  \centering
  \includegraphics[width=\linewidth]{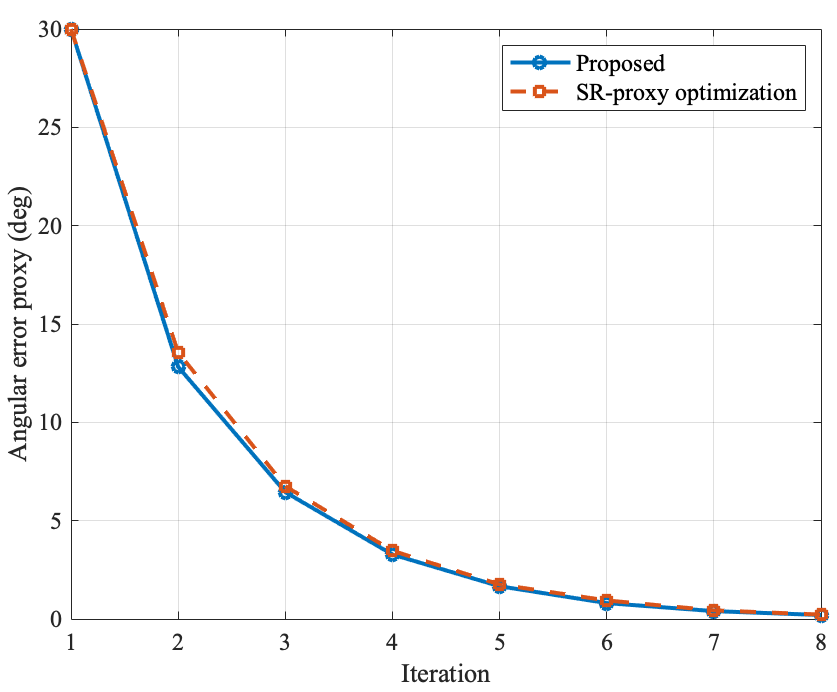}
  \caption{Sensing performance.}
  \label{fig:error}
\end{minipage}%
\hfill
\begin{minipage}[b]{0.32\linewidth}
  \centering
  \includegraphics[width=\linewidth]{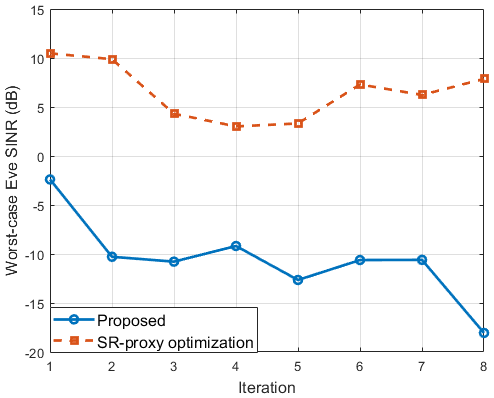}
  \caption{Security performance.}
  \label{fig:security}
\end{minipage}%
\hfill
\begin{minipage}[b]{0.32\linewidth}
  \centering
  \includegraphics[width=\linewidth]{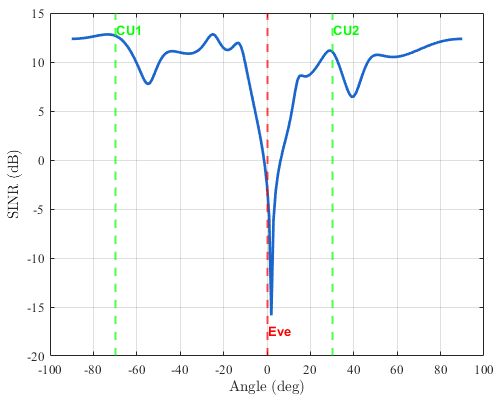}
  \caption{Final SINR beampattern.}
  \label{fig:beam}
\end{minipage}
\end{figure*}

Fig. \ref{fig:error}. illustrates the convergence of the Eve's localization error, defined as half of the angular uncertainty width.  It can be observed that both the proposed algorithm and SR-proxy optimization method effectively reduce the estimation error over iterations, thereby demonstrating their capability to enhance sensing performance.
Fig. \ref{fig:security}. presents the security performance, evaluated in terms of the maximum SINR at the Eve. The proposed algorithm successfully suppresses the maximum Eve SINR to a very low level, demonstrating its effectiveness in securing the communication link. 
In contrast, while the SR-proxy optimization method improves the secrecy rate for legitimate CUs, it fails to adequately suppress the maximum Eve SINR. 
This indicates a fundamental trade-off between communication and instantaneous security.

To provide a more intuitive illustration of the algorithm’s effectiveness, Fig. \ref{fig:beam}. presents the SINR beampattern across the angular domain. 
The final optimized beampattern clearly demonstrates the algorithm's ability to perform precise spatial power control. The high-gain beams are formed towards the directions of the legitimate CUs, ensuring their QoS requirements are satisfied. 
Simultaneously, a deep null is steered with high precision towards the Eve's final estimated location. This null significantly suppresses the signal power at the Eve, thereby effectively minimizing information leakage and guaranteeing communication security. 
Overall, these results demonstrate that the proposed algorithm can simultaneously enhance sensing precision and safeguard communication security, thereby providing a robust and practical solution for security-critical ISAC systems.

\section{Conclusion}
This paper addresses the challenge of ensuring instantaneous worst-case security in ISAC systems under unknown eavesdropper locations. Unlike conventional designs that rely on ergodic metrics or static robustness, we introduce a dynamic secure beamforming framework that co-designs sensing and security. A hierarchical iterative algorithm is developed, establishing an adaptive feedback loop between sensing-based uncertainty refinement and secure transmission design. By jointly minimizing information leakage and enhancing sensing accuracy, the framework achieves robust, per-transmission security guarantees in untrusted environments. Simulation results confirm its effectiveness, showing reduced localization error and significant suppression of worst-case Eve SINR.

Future work will focus on developing more refined performance metrics to capture the joint efficiency of communication, sensing, and security.
In addition, incorporating learning-based prior estimation could further enhance the adaptability of the proposed framework in highly dynamic environments, enabling robustness in practical deployments.

\section{Acknowledgement}
This research was supported in part, by the National Science and Technology Major Project of China (Project Number:2024ZD1300100), and the National Natural Science Foundation of China under Grant U24B20129 and 62401318.


\bibliographystyle{IEEEtran}
\bibliography{IEEEabrv,RefDefinition,ref}





\end{document}